\documentclass[conference]{IEEEtran}
\usepackage[T1]{fontenc}
\usepackage{graphicx,booktabs,amsmath,amssymb,xcolor,tikz}
\usepackage[hidelinks]{hyperref}
\usepackage{cite,xspace}
\usetikzlibrary{arrows.meta,positioning,decorations.pathreplacing}
\newcommand{\unit}[1]{\ifmmode\,\mathrm{#1}\else\,#1\xspace\fi}
\newcommand{\B}{\unit{B}}
\newcommand{\kB}{\unit{kB}}
\newcommand{\ms}{\unit{ms}}
\newcommand{\s}{\unit{s}}

\begin{document}
\title{Post-Quantum Session Establishment for Matter over Thread: Message Size, Transport, and Protocol Options}

\author{\IEEEauthorblockN{Daniel Commey}
\IEEEauthorblockA{TIDES Lab, Department of Computer Engineering and Computer Science\\
California State University, Long Beach, Long Beach, CA 90840, USA\\
daniel.commey@csulb.edu}}

\maketitle

\begin{abstract}
Matter's first post-quantum change protects device attestation with ML-DSA, but its operational handshake, CASE,
still derives session keys from P-256 ECDH, so recorded traffic can be decrypted once a quantum computer exists.
We quantify the cost of moving CASE to ML-KEM and post-quantum signatures over Thread, where frames carry at most
127 bytes and a Matter message at most 1178 bytes. A byte-exact model of the CASE messages shows that a drop-in
ML-KEM-768/ML-DSA-44 handshake grows from 1.6\kB to 22.8\kB and that no fully post-quantum configuration built
from FIPS~203/204/205 algorithms fits the message limit. A Monte Carlo model of multi-hop Thread, calibrated to
OpenThread and Matter source constants, shows that sending these messages as fragmented IPv6 datagrams causes
spurious retransmissions whenever a message and its reply outlast Matter's retransmission timer. At three to
four hops the drop-in handshake then takes 1.8 to 2.6 times longer than with a selective-repeat block transfer,
with or without interference, and up to 33\% of handshakes with sleepy devices fail; raising the timers does not
remove the effect. Caching certificate chains and authenticating the device with a pre-distributed ML-KEM key
reduce the handshake to 6.1\kB and 44\ms of device computation. As a first step, hybrid P-256+ML-KEM-512 key
exchange with a cached device chain protects session keys against harvest-now, decrypt-later attacks while every
message still fits the current limit.
\end{abstract}


\section{Introduction}
Matter~\cite{matterspec} is the IP-based application standard for smart-home devices. Battery-powered Matter
devices usually run over Thread, an IPv6 mesh on IEEE~802.15.4 radios with 127-byte frames and 250\,kb/s.
Matter devices authenticate each other with operational certificates, and every operational session starts with
Certificate Authenticated Session Establishment (CASE), a SIGMA-I handshake~\cite{sigma} over ephemeral P-256
ECDH with ECDSA-signed node operational certificates (NOCs).

A cryptographically relevant quantum computer would break both primitives. For key exchange the threat is
already present: an adversary who records CASE traffic today can recover the session keys later and decrypt
everything sent under them~\cite{mosca}. Smart-home traffic includes door-lock commands, occupancy, and camera
control, and devices stay in service for a decade or more, so the window is long. In September 2026 the Matter
reference implementation merged its first post-quantum change, \emph{PQC Phase~1}~\cite{chip-pqc-phase1}. It adds
ML-DSA to the device-attestation chain used during commissioning and carries the larger attestation certificates in
600-byte application-layer segments. CASE, the operational certificates, and the session keys are unchanged.

Migrating CASE is harder than swapping primitives because of Thread. ML-KEM-768 public keys are 1184\B and
ML-DSA-44 signatures 2420\B~\cite{fips203,fips204}. Matter caps a message at 1178\B so that it fits one
1280-byte IPv6 packet, and OpenThread disables IPv6 fragmentation by default~\cite{openthread}. A larger message
needs a new transport mechanism, and on a multi-hop 802.15.4 mesh the choice of mechanism interacts with Matter's
retransmission timers.

This paper quantifies those costs and evaluates protocol options. Our contributions are:
\begin{itemize}
\item A byte-exact size model of CASE for 15 representative configurations and 3780 design points, built on a
Matter TLV codec that round-trips all 19 certificate test vectors of the reference implementation
(\S\ref{sec:method}, \S\ref{sec:sizes}).
\item The result that no fully post-quantum CASE built from FIPS~203/204/205 algorithms fits a Matter message, so
migration requires a transport change, together with a configuration that gives harvest-now, decrypt-later
protection without one (\S\ref{sec:sizes}).
\item A Monte Carlo model of CASE over multi-hop Thread, with link, mesh, sleepy-device, and retransmission
parameters taken from OpenThread and Matter source code, and the identification of a timer pathology that makes
fragmented IPv6 transport 1.8 to 2.6 times slower than a block transfer at three or more hops
(\S\ref{sec:transport}).
\item Sensitivity analyses over interference burst length and occupancy, Matter retransmission timers, and segment
size (\S\ref{sec:sensitivity}), and a discussion of the security of the proposed protocol changes
(\S\ref{sec:discussion}).
\end{itemize}
All results are from models and simulation. We have not yet measured them on hardware; \S\ref{sec:limits} states
what this implies for each result.

\section{Background}\label{sec:background}

\subsection{CASE}
CASE has three messages and a status report. \emph{Sigma1} (initiator to responder) carries a 32-byte random,
a session identifier, a destination identifier derived from the target's fabric and node identity, the initiator's
ephemeral public key, and session parameters. \emph{Sigma2} returns the responder's ephemeral key and an encrypted
block containing the responder's NOC, an optional intermediate CA certificate (ICAC), a signature over both
ephemeral keys, and a resumption identifier. \emph{Sigma3} carries the initiator's encrypted NOC, ICAC, and
signature. The responder answers with a status report. Each party verifies the peer's chain up to the fabric's root
CA certificate (RCAC), which both store. Matter encodes certificates in a compact TLV form; the reference test
vectors have a 269-byte NOC and a 252-byte ICAC.

In a typical deployment the initiator is a controller (a phone, hub, or border router) and the responder is the
Thread device. Session resumption replaces Sigma2 and Sigma3 with a short Sigma2Resume authenticated by a key
derived from the previous session.

\subsection{Message size limit}
Over UDP, a Matter message payload may not exceed $1280 - 48 - 38 - 16 = 1178$\B: the IPv6 minimum MTU minus
uncompressed IPv6 and UDP headers, a reserve for Matter headers, and the message integrity code
(\texttt{kMaxAppMessageLen} in~\cite{connectedhomeip}). Matter defines a TCP transport for large payloads, but
Thread devices do not normally support it.

\subsection{Thread}
Thread carries IPv6 over 802.15.4 using 6LoWPAN header compression and fragmentation~\cite{rfc4944,rfc6282}.
With MAC security and short addresses, a 127-byte fragment frame carries 96 to 101\B of the IPv6 packet, so a
1232-byte UDP payload occupies 13 frames. Routers forward fragments hop by hop. OpenThread retransmits a unicast frame up
to 15 times, discards a partially reassembled packet after 2\s, and, for sleepy end devices (SEDs), holds frames
at the parent until the SED polls for them~\cite{openthread}.

\subsection{Matter reliable messaging}
The Message Reliability Protocol (MRP) acknowledges each reliable message, either piggybacked on the next
application message or, if none is sent within 200\ms, as a standalone acknowledgement. A sender retransmits the
whole message after
\begin{equation}
T_n = I\cdot 1.1\cdot 1.6^{\max(0,n-1)}\,(1+0.25\,U) + T_\mathrm{boost},\label{eq:mrp}
\end{equation}
where $I$ is the active retry interval advertised by the receiver, $n$ counts previous transmissions, and
$U\sim\mathcal{U}(0,1)$. A message is sent at most five times. On OpenThread builds a device advertises
$I=2000\ms$ and adds $T_\mathrm{boost}=1500\ms$ to its own timers; a controller on Wi-Fi or Ethernet advertises
$I=300\ms$~\cite{connectedhomeip}. The timer starts when the message is handed to the transport.

\section{Threat Model and Design Space}\label{sec:design}
We consider a passive adversary that records CASE exchanges now and later gains a quantum computer
(harvest-now, decrypt-later), and an active quantum adversary in the future. The first requires a post-quantum key
exchange in every CASE handshake performed today. The second requires post-quantum signatures, or another
post-quantum authentication mechanism, before such a computer exists. Authentication can therefore migrate
later than key exchange, but certificates issued now remain valid for years.

We vary five dimensions. \emph{Key exchange}: P-256 (today), ML-KEM-512 or ML-KEM-768~\cite{fips203}, or a hybrid
that concatenates a P-256 share with an ML-KEM encapsulation key, as in hybrid TLS~\cite{tlshybrid}.
\emph{Signature per chain position}: NOC, ICAC, and RCAC may each use ECDSA, ML-DSA-44~\cite{fips204},
Falcon-512, SLH-DSA-128s~\cite{fips205}, or one of the second-round additional signatures MAYO, SNOVA, or
UOV~\cite{nistaddsig}. CA keys sign offline, so slow signing is acceptable there; the NOC key signs during every
handshake. \emph{Chain structure}: with or without an ICAC. \emph{Responder authentication}: a signature as in
SIGMA; a static ML-KEM key in the responder's NOC, to which the initiator encapsulates in Sigma3 (KEM
authentication, as in KEMTLS~\cite{kemtls}); or, if the initiator already holds that key, encapsulation in Sigma1
and a key-confirmation MAC in Sigma2 (KEM-PDK, as in KEMTLS-PDK~\cite{kemtlspdk}). \emph{Cached chains}: if a party
already holds the peer's NOC and ICAC, the peer sends an 8-byte reference instead~\cite{rfc7924}. In Matter this
is common: the controller that commissioned a device issued its NOC.

Table~\ref{tab:configs} lists the 15 configurations used throughout. The design-space study in
\S\ref{sec:pareto} enumerates all 3780 combinations of 3 key exchanges, 5 NOC signatures, 6 ICAC options,
7 RCAC signatures, and 6 authentication and caching modes.

\begin{table*}[t]
\caption{Configurations. Chain lists NOC/ICAC/RCAC signature algorithms: E~=~ECDSA P-256, F~=~Falcon-512,
M~=~ML-DSA-44, Y~=~MAYO-2, U~=~UOV-Is-pkc. Cached: peer already holds the responder (R) or both (R+I) chains.
Payload sizes in bold exceed the 1178-byte limit; Total includes the status report. Device PQ: post-quantum
computation on the responder at 64\,MHz from pqm4 cycle counts (classical ECC excluded). Time: median
completion with block transfer, 1 hop without interference and 3 hops under heavy Wi-Fi interference.}
\label{tab:configs}
\centering\footnotesize
\setlength{\tabcolsep}{4.2pt}
\begin{tabular}{@{}lllllrrrrrrr@{}}
\toprule
& & & & & \multicolumn{4}{c}{Payload bytes} & Device & \multicolumn{2}{c}{Time [s]}\\
\cmidrule(lr){6-9}\cmidrule(l){11-12}
ID & KEM & Chain & Auth. & Cached & Sigma1 & Sigma2 & Sigma3 & Total & PQ [ms] & 1 hop & 3 hops\\
\midrule
B & P-256 & E/E/E & sig & -- & 174 & 775 & 619 & {1{,}576} & -- & 0.12 & 0.7\\
\multicolumn{12}{l}{\emph{Drop-in replacements}}\\
D1 & P-256+K768 & E/E/E & sig & -- & \textbf{1{,}359} & \textbf{1{,}864} & 619 & {3{,}850} & 10 & 0.29 & 1.6\\
D2 & P-256+K768 & F/F/F & sig & -- & \textbf{1{,}359} & \textbf{5{,}334} & \textbf{4{,}089} & {10{,}790} & 380 & 1.14 & 4.8\\
D3 & K768 & M/M/M & sig & -- & \textbf{1{,}294} & \textbf{11{,}361} & \textbf{10{,}181} & {22{,}844} & 139 & 1.71 & 9.3\\
D4 & P-256+K768 & E+M/E+M/E+M & sig & -- & \textbf{1{,}359} & \textbf{11{,}760} & \textbf{10{,}515} & {23{,}642} & 139 & 1.76 & 9.7\\
\multicolumn{12}{l}{\emph{Structural changes}}\\
S1 & K768 & M/--/M & sig & -- & \textbf{1{,}294} & \textbf{7{,}503} & \textbf{6{,}323} & {15{,}128} & 116 & 1.17 & 6.3\\
S2 & K768 & M/--/Y & sig & -- & \textbf{1{,}294} & \textbf{5{,}268} & \textbf{4{,}088} & {10{,}658} & 188 & 0.94 & 4.6\\
S3 & K768 & M/--/U & sig & -- & \textbf{1{,}294} & \textbf{5{,}178} & \textbf{3{,}998} & {10{,}478} & 94 & 0.83 & 4.4\\
S4 & K768 & M/M/M & KEM & -- & \textbf{1{,}294} & \textbf{8{,}809} & \textbf{11{,}273} & {21{,}403} & 88 & 1.56 & 8.7\\
S5 & K768 & M/--/M & KEM & -- & \textbf{1{,}294} & \textbf{4{,}951} & \textbf{7{,}415} & {13{,}687} & 66 & 1.02 & 5.7\\
\multicolumn{12}{l}{\emph{Prior knowledge of peer credentials}}\\
C1 & P-256+K512 & E/E/E & sig & R & 975 & 1{,}026 & 619 & {2{,}628} & 6 & 0.19 & 1.0\\
C2 & K768 & M/M/M & sig & R & \textbf{1{,}294} & \textbf{3{,}639} & \textbf{10{,}181} & {15{,}122} & 139 & 1.19 & 6.2\\
C3 & K768 & M/M/M & sig & R+I & \textbf{1{,}294} & \textbf{3{,}639} & \textbf{2{,}459} & {7{,}400} & 94 & 0.63 & 3.1\\
C4 & K768 & M/--/M & KEM-PDK & -- & \textbf{2{,}386} & \textbf{1{,}222} & \textbf{6{,}323} & {9{,}939} & 66 & 0.77 & 4.1\\
C5 & K768 & M/--/M & KEM-PDK & R+I & \textbf{2{,}386} & \textbf{1{,}222} & \textbf{2{,}459} & {6{,}075} & 44 & 0.49 & 2.6\\
\bottomrule
\end{tabular}

\end{table*}

\section{Methodology}\label{sec:method}

\subsection{Message size model}
We implemented the Matter TLV encoding and decode the certificate test vectors of the reference
implementation~\cite{connectedhomeip}; all 19 re-encode to identical bytes. Post-quantum certificates are built
from the real NOC, ICAC, and RCAC templates by replacing the public key and signature fields and dropping the
elliptic-curve identifier; all other fields, including extensions, are unchanged. A hybrid (composite) certificate
carries its second key and signature under two new context tags. Matter has not specified post-quantum operational
certificates, so these encodings are our assumption; they add only TLV length bytes beyond the raw key and
signature sizes. The CASE messages follow the field layout of \texttt{CASESession.cpp}, including session
parameters. Fields we introduce (KEM ciphertexts, certificate references, key-confirmation MACs) use new
context tags. Algorithm sizes are from the standards and were checked against liboqs~0.16.

\subsection{Computation model}
Device computation counts the post-quantum operations on the responder: encapsulation and signing before Sigma2,
decapsulation and verification after Sigma3. We use the fastest Cortex-M4 cycle counts in pqm4~\cite{pqm4}
(measured at 24\,MHz with zero wait states) and scale to 64\,MHz, the clock of the nRF52840. We exclude classical
ECC from all configurations, because its cost depends on the hardware accelerator, so the classical baseline is
slightly optimistic. The controller is assigned 5\ms per message.

\subsection{Transport simulation}
The simulator follows each 802.15.4 frame on each hop. Per frame it models the unslotted CSMA-CA initial backoff,
airtime at 32\,$\mu$s per byte including the synchronization header, turnaround, the immediate acknowledgement or
acknowledgement wait, and up to 15 retransmissions with OpenThread's retransmission backoff. Frames carry 6LoWPAN
fragmentation and, beyond one hop, a mesh header; a datagram is lost if any fragment exhausts its retries or
reassembly exceeds 2\s. For SEDs the parent delivers each frame after a data poll; a frame-pending bit lets the
device poll again immediately for consecutive frames, and each frame has four indirect attempts. Hops are
traversed store-and-forward on a single shared channel without spatial reuse. This lengthens multi-hop
transfers relative to a pipelined mesh and therefore favors the timer effect in \S\ref{sec:transport}; we return to
this point there. MRP is modeled with Eq.~\eqref{eq:mrp}, five transmissions, piggybacked and standalone
acknowledgements, and duplicate transmissions that occupy the channel. Table~\ref{tab:params} lists the
parameters and their sources.

Channel loss follows a time-based Gilbert--Elliott model~\cite{gilbert,elliott}: the channel alternates between
a good state with loss probability $p_g$ and a bad state with loss probability 0.9, with exponentially distributed
durations. This reproduces the bursty loss that 802.15.4 links experience under Wi-Fi
interference~\cite{liang2010,betafactor}. We use four scenarios: no interference; light Wi-Fi ($p_g=1\%$, 10\% of
time in 10-ms bursts); heavy Wi-Fi ($p_g=2\%$, 30\% in 30-ms bursts); and an i.i.d.\ lossy link ($p_g=15\%$).
These values are illustrative rather than measured, and \S\ref{sec:sensitivity} varies them.

\begin{table}[t]
\caption{Model parameters. Source: connectedhomeip @8982f4d (CHIP), OpenThread @75bf129 (OT).}
\label{tab:params}
\centering\footnotesize
\setlength{\tabcolsep}{3pt}
\begin{tabular}{@{}llr@{}}
\toprule
Parameter & Source & Value\\
\midrule
Max.\ Matter payload (\texttt{kMaxAppMessageLen}) & CHIP & 1178\B\\
IPv6 MTU, IPv6 fragmentation & OT & 1280\B, off\\
MAC header + security + FCS & 802.15.4 & 21\B\\
Direct frame retransmissions & OT & 15\\
Indirect (SED) attempts per frame & OT & 4\\
6LoWPAN reassembly timeout & OT & 2\s\\
MRP active interval, device / controller & CHIP & 2000 / 300\ms\\
MRP sender boost, Thread device & CHIP & 1500\ms\\
MRP standalone-ACK timeout & CHIP & 200\ms\\
MRP transmissions & CHIP & 5\\
SED poll interval during CASE & assumed & 200\ms\\
Device clock / controller per message & assumed & 64\,MHz / 5\ms\\
\bottomrule
\end{tabular}
\end{table}

\emph{Transports.} Messages that fit are sent as one datagram. For larger messages we compare three mechanisms.
\emph{IPv6 fragmentation} sends the message as $\lceil s/1232\rceil$ IPv6 datagrams; this requires enabling IPv6
fragmentation, and a lost datagram loses the message, which MRP then resends in full. \emph{Pull segmentation}
follows PQC Phase~1: the receiver requests each segment, and each request--segment pair is an MRP exchange.
\emph{Block transfer} is our proposal, similar in spirit to CoAP's Q-Block option~\cite{rfc9177}: the sender
transmits all segments back to back, the receiver returns a bitmap of received segments, and the sender resends
only the missing ones. If the bitmap is lost, the sender retransmits unconfirmed segments after the MRP timer.
Segments carry 1024 payload bytes and a 12-byte header unless stated otherwise.

\emph{Metrics.} Handshake time runs from the controller sending Sigma1 until the status report reaches the
controller, when both parties hold the session; it excludes the 200-ms wait before the final standalone
acknowledgement. We also record failures (any message exhausting MRP), retransmissions, and the device's radio
transmit and receive time. Each scenario is run 300 times with independent seeds; we report medians with
percentile-bootstrap 95\% confidence intervals and the 90th percentile. The full grid has 15 configurations,
1 to 4 hops, 4 interference scenarios, always-on and sleepy devices, and 3 transports, for 432{,}000 handshakes.

\emph{Piggybacked acknowledgements.} The arrival time of a piggybacked acknowledgement is estimated as the
loss-free delivery time of the first unit (whole message, or first segment) of the reply. Under loss the real
arrival is later, so this approximation \emph{underestimates} spurious retransmissions, which works against our
finding in \S\ref{sec:transport}.

\section{Message Sizes}\label{sec:sizes}

\begin{figure}[t]\centering
\includegraphics[width=\columnwidth]{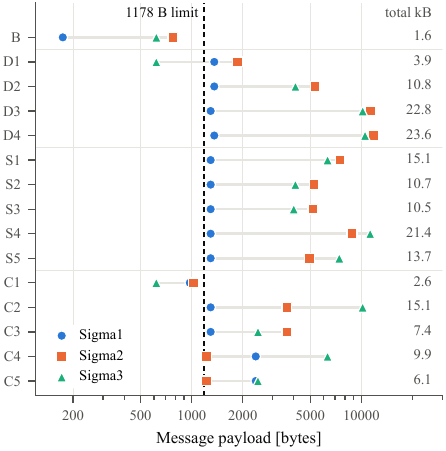}
\caption{Payload of each Sigma message per configuration (Table~\ref{tab:configs}), log scale. Messages right of
the dashed line exceed Matter's 1178-byte limit; numbers on the right give the handshake total.}
\label{fig:sizes}
\end{figure}

Fig.~\ref{fig:sizes} and Table~\ref{tab:configs} give the message sizes. Today's handshake (B) is 1576\B in 20
frames at one hop, and its largest message, Sigma2, is 775\B. Adding ML-KEM-768 to the key exchange (D1) already
breaks the limit twice: the hybrid share makes Sigma1 1359\B, and the ciphertext makes Sigma2 1864\B. A drop-in
post-quantum handshake with ML-KEM-768 and an all-ML-DSA-44 chain (D3) is 22.8\kB in 244 frames, 14 times the
classical size. Composite hybrid certificates (D4) add another 0.8\kB.

Structural changes help but do not reach the limit. Removing the ICAC (S1) saves 7.7\kB. A short-signature root
(MAYO-2, S2, or UOV, S3) shrinks each NOC because the NOC carries the issuer's signature, but places a 4.9\kB or
66\kB root key in device storage. KEM authentication replaces the responder's 2420-byte signature with a 1088-byte
ciphertext (S4, S5) but moves the ciphertext into Sigma3.

Prior knowledge of credentials has the largest effect. Caching both chains (C3) removes 15.4\kB from D3. With
the device's ML-KEM key pre-distributed and both chains cached (C5), the handshake is 6.1\kB, and the device
performs one encapsulation, one decapsulation, and one ML-DSA verification: 44\ms instead of 139\ms. Its largest
message, Sigma3 at 2459\B, is dominated by the controller's ML-DSA-44 signature.

\emph{No fully post-quantum configuration fits.} Across all 3780 design points, the smallest largest message using
only FIPS~203/204/205 algorithms is 2459\B (ML-KEM with KEM-PDK and both chains cached). Allowing Falcon-512,
whose standard (FIPS~206) is not final, lowers it to 1565\B. The only points where every message fits use
ML-KEM-512, both chains cached, and a MAYO or SNOVA NOC key, schemes still under evaluation. A post-quantum
handshake therefore needs a way to carry messages above 1178\B.

\emph{Key exchange alone can fit.} The harvest-now, decrypt-later threat only requires a post-quantum key
exchange. A hybrid P-256+ML-KEM-512 share with classical certificates and a cached responder chain (C1) yields
messages of 975, 1026, and 619\B, all within the limit, for a 2.6\kB handshake. ML-KEM-512 targets NIST security
category~1, comparable to the 128-bit classical security of P-256, and the hybrid keeps P-256. C1 needs only a
certificate-reference field in Sigma2 and the hybrid share; it does not need a new transport. Some national
guidance recommends ML-KEM-768 or stronger for general use. With ML-KEM-768 the cached-chain Sigma2 is 1346\B and
does not fit, so a deployment that follows that guidance needs the transport changes of \S\ref{sec:transport} even
for key exchange alone.

\emph{Resumption.} The reference implementation requires the ephemeral key in every Sigma1, including resumption
attempts. With an ML-KEM-768 share this makes a resumption Sigma1 1332\B, above the limit. The share is used
only if resumption fails and the responder falls back to a full Sigma2 (\texttt{CASESession.cpp} returns before
reading it when resumption succeeds). Making it optional reduces a post-quantum resumption to 226\B in total,
below today's 294\B, at the cost of an extra round trip when resumption fails.

\section{Transport for Large Messages}\label{sec:transport}

\begin{figure*}[t]\centering
\includegraphics[width=\textwidth]{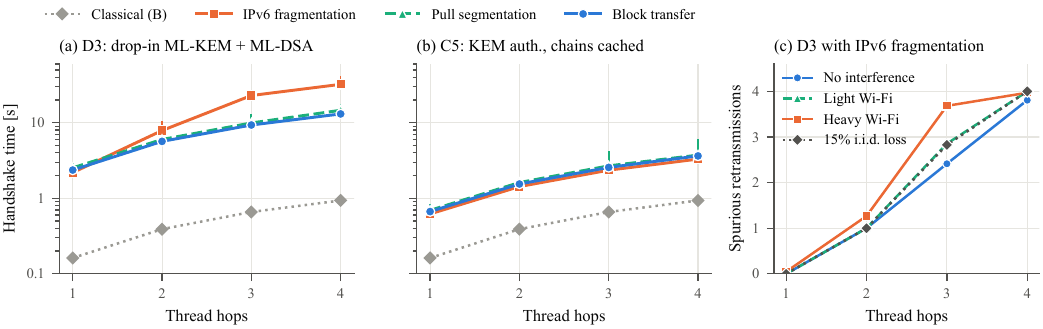}
\caption{Handshake time under heavy Wi-Fi interference for (a) the drop-in configuration D3 and (b) the smallest
post-quantum configuration C5, by transport. Markers show medians (95\% bootstrap confidence intervals are smaller
than the markers); vertical bars extend to the 90th percentile. (c) Mean spurious retransmissions per D3 handshake
with IPv6 fragmentation, by interference scenario. Always-on device, 300 runs per point.}
\label{fig:transport}
\end{figure*}

\begin{figure}[t]\centering
\begin{tikzpicture}[font=\footnotesize, >={Stealth[length=4pt]}, x=0.88cm, y=0.6cm]
\draw[thick] (0,0) node[left] {Device} -- (8.2,0);
\draw[thick] (0,-3) node[left] {Controller} -- (8.2,-3);
\draw[->, thick] (0.2,0) -- (2.6,-3);
\node[anchor=east] at (1.25,-1.3) {Sigma2};
\draw[->, thick] (2.7,-3) -- (6.6,0);
\node[anchor=west] at (3.95,-2.6) {Sigma3 + ACK};
\draw[->, thick, dashed] (4.3,0) -- (7.0,-3);
\node[anchor=north] at (7.0,-3.1) {Sigma2 resent};
\draw[decorate, decoration={brace, amplitude=3pt}] (0.2,0.35) -- node[above=2pt] {MRP timer $T_0$} (4.3,0.35);
\fill (4.3,0) circle (1.8pt);
\fill (6.6,0) circle (1.8pt);
\node[anchor=south, align=center] at (6.6,0.15) {ACK received};
\end{tikzpicture}
\caption{Spurious retransmission with IPv6 fragmentation. The device's MRP timer for Sigma2 starts when Sigma2 is
handed to the network and expires before the acknowledgement, piggybacked on the fully reassembled Sigma3,
arrives. The device resends all of Sigma2, which competes for the channel with Sigma3.}
\label{fig:pathology}
\end{figure}
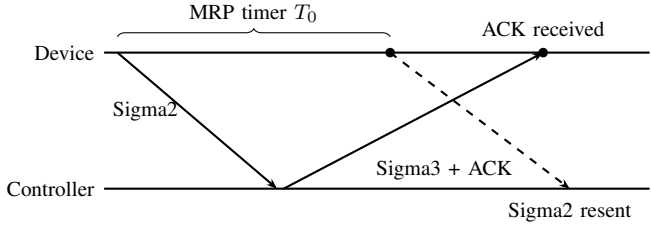

Fig.~\ref{fig:transport} compares the three transports. At one hop, IPv6 fragmentation is the fastest: for D3 it
completes in a median 1.58\s without interference, against 1.71\s for block transfer and 1.82\s for pull
segmentation, because it sends no segment headers or bitmaps. From two hops on it is the slowest. At three hops
without interference D3 takes 8.8\s with IPv6 fragmentation and 4.9\s with block transfer; at four hops under heavy
interference, 32.3\s and 13.1\s.

The cause is the interaction shown in Fig.~\ref{fig:pathology}. The device sends Sigma2 and starts a timer of
$T_0 = 300\cdot 1.1\cdot(1+0.25U) + 1500 \approx 1.83$--$1.91\s$. The acknowledgement rides on Sigma3, and with IPv6
fragmentation the controller's stack delivers Sigma3, and with it the acknowledgement, only after the last
fragment arrives. A spurious retransmission occurs when
\begin{equation}
T_\mathrm{tx}(\text{Sigma2}) + t_\mathrm{ctrl} + T_\mathrm{tx}(\text{Sigma3}) > T_0 .
\end{equation}
Loss-free, Sigma2 and Sigma3 of D3 need about 0.71\s and 0.64\s per hop on a mesh path, so the condition holds from two hops on
even without loss. The retransmitted 11-kB Sigma2 then delays Sigma3 further, and the next timer, 1.6 times
longer, can expire again. Fig.~\ref{fig:transport}c shows 1.0 spurious retransmissions per handshake at two hops
and 3.8 at four hops without interference, nearly the MRP maximum of four. Interference adds
losses, but the retransmissions themselves are caused by timing. The same holds in the other direction: Sigma3 travels with
the controller's timer of at least 2.2\s, which it can also exceed at four hops.

With pull segmentation and block transfer the acknowledgement for Sigma2 is piggybacked on the first segment of
Sigma3, so no message waits for a large reply. Across the 416 scenarios where a segmented transport is used, pull
segmentation has a median handshake time 5.9\% longer than block transfer (range $-4.5\%$ to $+37\%$); it was
faster in 9 scenarios. The largest differences occur with SEDs, where every segment request waits for a poll.

For the smallest configuration C5 the largest message is 2.4\kB, the timer condition never holds, and IPv6
fragmentation is 7 to 9\% faster than block transfer at all hop counts (Fig.~\ref{fig:transport}b). Spurious
retransmissions are thus specific to large handshakes.

\emph{Sleepy devices.} SEDs receive each frame only after polling. With a 200-ms poll interval and heavy
interference, IPv6 fragmentation lost handshakes in 28 of the 52 sleepy-device scenarios with oversized
messages, up to 33\% of them (S4 at four hops) and 24\% for D3. All of these failures occurred in Sigma3, the
large message toward the device: each 1232-byte datagram needs 13 consecutive indirect frames, and four failed
attempts on any of them discard the whole message. (Two always-on scenarios at four hops also lost one handshake
each.) Over the oversized configurations, block transfer completed all 124{,}800 handshakes it was used for, and
pull segmentation all but one.

\emph{Radio time.} At one hop without interference the device's radio transmits or receives for 85\ms in a
classical handshake, 1.15\s for D3, 318\ms for C5, and 133\ms for C1. For sleepy devices, whose radio is
otherwise off, handshake energy scales roughly with this time, so caching and KEM authentication reduce it in
proportion to the handshake size. Always-on Thread routers keep their receiver on regardless, so for them the
saving is smaller.

\section{Sensitivity}\label{sec:sensitivity}

\begin{figure*}[t]\centering
\includegraphics[width=\textwidth]{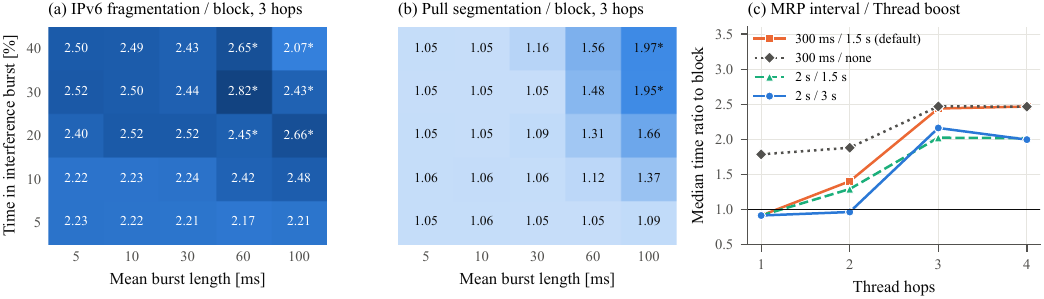}
\caption{Sensitivity of the transport comparison for D3. (a, b) Ratio of median handshake time to block transfer at
three hops as a function of interference occupancy and mean burst length; an asterisk marks success below 99.5\%.
(c) Ratio of IPv6 fragmentation to block transfer under heavy interference for combinations of the controller's
advertised active interval and the Thread device's sender boost. 200 runs per cell.}
\label{fig:sensitivity}
\end{figure*}

\emph{Interference model.} Fig.~\ref{fig:sensitivity}a varies the fraction of time in a bad state from 5\% to
40\% and the mean burst length from 5 to 100\ms at three hops. IPv6 fragmentation is 2.1 to 2.8 times slower than
block transfer in all 25 settings, and with bursts of 60\ms or more it also loses up to 98\% of handshakes. Pull
segmentation (Fig.~\ref{fig:sensitivity}b) stays within 16\% of block transfer for bursts up to 30\ms but becomes
1.3 to 2.0 times slower when bursts of 60 to 100\ms occupy 20\% or more of the time. A long burst destroys several consecutive segments; block
transfer reports all of them in one bitmap, while pull segmentation recovers each with a separate MRP timeout of at
least 2.2\s.

\emph{Retransmission timers.} An implementer could raise the timers to avoid spurious retransmissions.
Fig.~\ref{fig:sensitivity}c varies the controller's advertised interval (300\ms to 2\s) and the device's boost
(none to 3\s). A 2-s interval with a 3-s boost removes the effect at two hops, but at three and four hops IPv6
fragmentation remains 2.0 to 2.5 times slower than block transfer for all nine settings we tested. Longer timers
also delay recovery from real losses.

\begin{figure}[t]\centering
\includegraphics[width=\columnwidth]{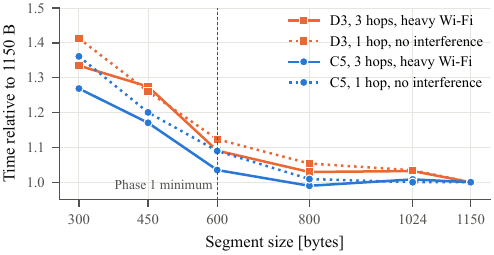}
\caption{Median handshake time with pull segmentation, relative to 1150-byte segments, as a function of segment
size. 200 runs per point.}
\label{fig:segments}
\end{figure}

\emph{Segment size.} PQC Phase~1 allows segments down to 600\B. Fig.~\ref{fig:segments} shows that segment
size matters more than the choice between pull and block. Relative to 1150-byte segments, 300-byte segments are
27 to 41\% slower and 600-byte segments 3 to 12\% slower, because every segment adds a request, an MRP exchange,
and a partially filled final frame. Segments should be as large as the 1178-byte limit allows.

\section{Design-Space Trade-offs}\label{sec:pareto}

\begin{figure}[t]\centering
\includegraphics[width=\columnwidth]{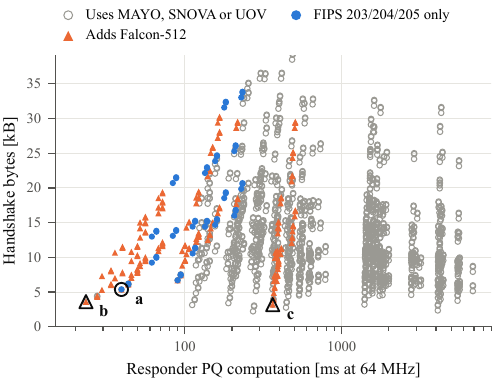}
\caption{Total handshake bytes against post-quantum computation on the device for 3780 design points. Black
outlines mark the Pareto-optimal points: (a) C5 with ML-KEM-512, the only one using FIPS~203/204/205 algorithms
alone (5.4\kB, 39\ms); with Falcon-512 allowed, (b) KEM-PDK with a Falcon-512 key in the controller's NOC and both chains cached (3.6\kB, 23\ms) and
(c) signature-based with Falcon-512 NOCs and both chains cached (3.2\kB, 363\ms). Points with identical size and
computation overlap.}
\label{fig:pareto}
\end{figure}

Fig.~\ref{fig:pareto} plots handshake size against device computation. With FIPS~203/204/205 algorithms only, a
single configuration is Pareto-optimal: C5 with ML-KEM-512 (5.4\kB, 39\ms). With Falcon-512 two more appear: a
KEM-PDK variant in which the controller's NOC carries a Falcon-512 key (3.6\kB, 23\ms) and a signature-based one with Falcon NOCs
(3.2\kB, 363\ms). In the first, only the controller signs, so the device merely verifies Falcon signatures.
In the second the device signs with Falcon, which needs double-precision floating point that the Cortex-M4F
lacks and must emulate; it trades 0.4\kB for about 340\ms of device computation and depends on FIPS~206, which
is not yet final. Most points with MAYO, SNOVA, or
UOV are dominated: they shorten signatures but cost seconds of signing or verification on the device, or very
large public keys.

\section{Discussion}\label{sec:discussion}

\subsection{Security of the protocol options}
We give an informal analysis; a formal treatment is future work.

\emph{Hybrid key exchange.} Concatenating P-256 and ML-KEM shares and deriving the session key from both
secrets keeps the session secure if either problem remains hard~\cite{tlshybrid}. C1 keeps classical
authentication, so it protects against recording now and decrypting later, but not against an active quantum
adversary.

\emph{KEM authentication.} In S4 and S5 the responder proves possession of its static KEM key by decapsulating the
initiator's ciphertext and confirming the derived key, as in KEMTLS~\cite{kemtls}. The ephemeral ML-KEM exchange
still provides forward secrecy. Responder authentication becomes implicit and is confirmed one message later,
and this weakens identity protection. SIGMA-I reveals the initiator's identity only after the responder has
authenticated. In S4 and S5, Sigma2 carries no signature, so an active attacker can relay the device's public
certificate, complete the ephemeral exchange itself, and decrypt Sigma3, which contains the controller's
certificate. The attacker cannot complete the handshake, but it learns who the initiator is. With KEM-PDK (C4, C5) the initiator encapsulates to the responder's known key in Sigma1, so only the intended
responder can complete Sigma2, and the initiator verifies Sigma2's key confirmation before sending its own
certificate in Sigma3, which preserves SIGMA-I's identity protection. The initiator must obtain the responder's key
authentically, which in Matter is natural because the controller issued the device's NOC. We therefore recommend
KEM-PDK rather than plain KEM authentication.

\emph{Certificate references.} A reference is sent inside the encrypted part of Sigma2 or Sigma3, so it reveals no
more about identity than the certificate it replaces. If the receiver lacks the referenced chain, the handshake must
fall back to sending it, which costs a round trip but no security.

\emph{Resumption without a KEM share.} The resumption keys derive from the previous session's shared secret and fresh
randoms, so omitting the share does not change the security of a successful resumption. As today, resumption does
not add forward secrecy beyond the original session.

\subsection{Recommendations}
Our results suggest a staged migration. (1)~Deploy hybrid P-256+ML-KEM-512 with a certificate reference for the
device chain (C1): it closes the harvest-now, decrypt-later gap without a new transport and costs 70\ms at one hop.
Where policy requires ML-KEM-768, this step needs the transport of step~(2).
(2)~Specify a block transfer for secure-channel messages instead of relying on IPv6 fragmentation, and reuse it
for the attestation certificates of PQC Phase~1. Segments should be as large as the message limit allows, and the
number of segments in flight should be bounded by a window so that senders and routers with small buffer pools are
not overrun.
(3)~Specify post-quantum NOCs together with chain references and KEM-PDK device authentication, which keeps a
fully post-quantum handshake near 6\kB and the device's computation below 50\ms. (4)~Make the ephemeral share in
resumption Sigma1 optional.

\subsection{Limitations}\label{sec:limits}
All results come from models. The message sizes are exact for the stated encodings, but post-quantum certificate
formats are not yet specified. The latency results depend on the simulator: it serializes transmissions on one
channel, ignores spatial reuse and other traffic, and estimates piggybacked acknowledgement times without loss.
Ignoring spatial reuse overstates multi-hop transfer times, which favors the timer effect at three or more hops;
at two hops, where neighboring links interfere and pipelining gains little, the effect already appears without
loss. The simulator also does not model memory. OpenThread's default build has 44 message buffers of 128\B,
about 5.6\kB, which cannot hold a 10-kB Sigma message at once. Larger pools are a build option, but IPv6
fragmentation and an unwindowed block transfer both need the whole message buffered, whereas pull segmentation
needs one segment, so default buffer sizes may favor pull segmentation. We have not evaluated Matter's TCP
transport: Thread devices rarely implement it, and it adds a connection setup and per-connection state before
CASE begins, but TCP over Thread can perform well~\cite{tcplp} and is a natural baseline for future work.
Absolute times should therefore be read as estimates; the ratios between transports and their sensitivity to
parameters are the more robust results, since they hold across all settings in \S\ref{sec:sensitivity}. Device
computation uses Cortex-M4 cycle counts and excludes classical ECC; other cores (for example RISC-V) and hardware
accelerators will differ. The interference scenarios are illustrative. Validating frame counts, latency, and
energy on nRF52840 and ESP32-H2 devices with a sniffer and power analyzer is our next step.

\section{Related Work}\label{sec:related}
\emph{Post-quantum Matter.} PQC Phase~1~\cite{chip-pqc-phase1} is, to our knowledge, the only post-quantum work in
Matter, and it covers device attestation only. Measurement studies of Matter over Thread, such as~\cite{nobile2026},
address application performance rather than security handshakes.

\emph{Constrained key exchange.} EDHOC~\cite{rfc9528} is the IETF's lightweight authenticated key exchange for
constrained devices. Post-quantum cipher suites for it~\cite{lakepq} and a KEM-based authentication
method~\cite{edhocauthkem} have been proposed; neither studies the transport of large messages over an 802.15.4
mesh. Our results on message size and transport apply to them as well.

\emph{Post-quantum TLS.} KEMTLS~\cite{kemtls} and KEMTLS-PDK~\cite{kemtlspdk} replace handshake signatures with KEM
operations, and we adapt both to CASE. Sikeridis et al.~\cite{sikeridis2020} and Paquin et al.~\cite{paquin2020}
measured how post-quantum signatures and key exchange affect TLS latency under packet loss and found that larger
handshakes suffer disproportionately on lossy links, consistent with our results.
B\"urstinghaus-Steinbach et al.~\cite{burstinghaus2020} integrated Kyber and SPHINCS+ into mbed TLS on embedded
platforms. Certificate caching~\cite{rfc7924} and compression~\cite{rfc8879} reduce TLS handshake size.
Our earlier benchmark~\cite{commey2025pqcce} measured post-quantum primitives on consumer-electronics platforms.

\emph{Large transfers on low-power networks.} CoAP block-wise transfer~\cite{rfc7959} and its Q-Block
extension~\cite{rfc9177} move large payloads over lossy constrained networks, the latter with selective
retransmission similar to our block transfer. TCPlp~\cite{tcplp} showed that TCP can perform well over Thread.

\section{Conclusion}\label{sec:conclusion}
The main cost of moving Matter's CASE handshake to post-quantum cryptography is communication on Thread;
device computation is secondary. A drop-in ML-KEM/ML-DSA handshake is 14 times larger than today's, and no fully post-quantum
configuration using FIPS~203/204/205 algorithms fits Matter's 1178-byte message. Simply allowing fragmented IPv6
datagrams is not an adequate fix on multi-hop networks: large messages and their replies outlast Matter's
retransmission timers, causing spurious retransmissions that make the handshake up to 2.6 times slower, and
sleepy devices fail up to a third of handshakes. A block transfer avoids both effects. Caching chains and
authenticating the device with a pre-distributed ML-KEM key reduce a post-quantum handshake to 6.1\kB, and a hybrid
ML-KEM-512 key exchange with a cached device chain protects against recorded traffic today without changing the
transport. Hardware measurements will test the latency estimates and quantify energy.

\section*{Artifact Availability}
The TLV codec, size model, simulator, and the scripts that produce every table and figure in this paper will be
released as open source.

\bibliographystyle{IEEEtran}
\bibliography{refs}
\end{document}